\documentclass[]{elsarticle}
\usepackage[]{graphicx}
\usepackage{amssymb}
\usepackage[english]{babel}

\usepackage{color}

\newcommand{\RV}[1]{{\color{red}{#1}}}

\def\mug{$\mu$g{} }
\newcommand{\DEL}[1]{}   %    	{{\color{blue}\sout{#1}}}

\begin{document}

%%%%%%%%%%%%%%%%%%%%%%%%%%%%%%%%%%%%%%%%%%%%%%%%%%%%%%%%%%%%%%%%%%%%%%%%%%%%%
%% Frontmatter
\begin{frontmatter}

\title{Realization of hydrodynamic experiments on
quasi-$2$D liquid crystal films in microgravity}

\author{Noel A.~Clark$^1$, Alexey Eremin$^2$, Matthew A.~Glaser$^1$, Nancy Hall$^3$, Kirsten Harth$^2$, Christoph Klopp$^2$, Joseph E.~Maclennan$^1$, Cheol S.~Park$^1$, Ralf Stannarius$^2$, Padetha Tin$^4$, William N. Thurmes$^5$, Torsten Trittel$^2$}
\address{
$^1$Department of Physics, University of Colorado Boulder, Boulder, CO, USA\\
$^2$Institute of Experimental Physics, Otto von Guericke University, Magdeburg, Germany \\
$^3$ NASA Glenn Research Center, Cleveland, OH, USA\\
$^4$ Universities Space Research Association, Cleveland, OH, USA\\
$^5$ Miyota Development Center of America, Longmont, CO, USA}

\cortext[cor1]{Corresponding authors: R.S., ralf.stannarius@ovgu.de, J.E.M., jem@colorado.edu
}

%\ead{ralf.stannarius@ovgu.de,jem@colorado.edu}

\begin{abstract}
Freely suspended films of smectic liquid crystals are unique examples of quasi two-dimensional fluids. Mechanically stable and with
quantized thickness of the order of only a few molecular layers, smectic films are ideal systems for studying fundamental fluid physics,
such as collective molecular ordering, defect and fluctuation phenomena, hydrodynamics, and nonequilibrium behavior in two dimensions
(2D), including serving as models of complex biological membranes. Smectic films can be drawn across openings in planar supports
resulting in thin, meniscus-bounded membranes, and can also be prepared as bubbles, either supported on an inflation tube or floating
freely.
%{These geometries are available for long-term studies by the very low liquid crystal vapor pressure.}
The quantized layering renders
%\DEL{properties like density and viscosity uniform to degree comparable to that of bulk fluids, making thin}
smectic films uniquely useful in 2D fluid physics. The OASIS team has pursued a variety of
ground-based and microgravity applications of thin liquid crystal films to fluid structure and hydrodynamic problems in 2D and
quasi-2D systems. Parabolic flights and sounding rocket experiments were carried out in order to explore the shape evolution of free
floating smectic bubbles, and to probe Marangoni effects in flat films. The dynamics of emulsions of smectic islands (thicker regions
on thin background films) and of microdroplet inclusions in spherical films, as well as thermocapillary effects, were studied over extended
periods within the OASIS (Observation and Analysis of Smectic Islands in Space) project on the International Space Station. We summarize
the technical details of the OASIS hardware and give preliminary examples of key observations.
\end{abstract}

\begin{keyword}
liquid crystal \sep smectic \sep microgravity \sep $2$D hydrodynamics
 \PACS 05.45.-A\sep 45.50.-j\sep 81.70.Ha\sep 83.10.Pp
\end{keyword}

\end{frontmatter}

\parindent=0.5 cm

%%%%%%%%%%%%%%%%%%%%%%%%%%%%%%%%%%%%%%%%%%%%%%%%%%%%%%%%%%%%%%%%%%%%%%%%%%%%%
%% Main text

\section{Introduction}

\subsection{Smectic freely suspended films}
Smectic liquid crystals are layered phases of rod-shaped organic molecules, in which each layer is a two-dimensional fluid on the order of a molecular length in  thickness.  Because of their layering, smectics can readily be made to form films that are freely suspended in air or vacuum.  These fluid structures are quantized in thickness, everywhere corresponding to some integral number $N$ of smectic layers, which can be as small as a single molecular monolayer  ($\approx 3$~nm  thick) for some materials.

Smectic films are structures of fundamental interest in  condensed matter physics. They are the thinnest known stable structures of any fluid phase preparation and have the largest surface-to-volume ratio, making them ideal for
studies of fluctuations, interface phenomena, and the
effects of reduced dimensionality in soft matter. The layering
makes films of uniform layer number homogeneous in
basic physical properties such as thickness, density, surface
tension, and viscosity, to a degree comparable to that of
bulk 3D fluids. This distinguishes them from soap films,
for example, which generally vary in thickness. Freely suspended
smectic films are, except at their edges, completely
free of local pinning or other external spatial inhomogeneities.
The typically low liquid crystal vapor pressure
and the absence of solvent enables the potential for convenient
long-term study of such films. The interactions which
are operative in ultra-thin films are generally weak, leading
to their easy manipulation by external agents such as
boundaries, applied fields, and flows, as well as enabling
significant fluctuation phenomena with extended spatial
correlations. These fluctuation, field, and surface effects,
combined with the wide variety of liquid crystal order
parameters and symmetries, make freely suspended LC
films a rich system for probing basic fluid physics.

\subsection{Fundamental research on thin smectic films}

The initial applications of ultra-thin freely suspended
liquid crystals were studies of the structural and fluctuation
behavior of smectic C phases, mesophases in which local
molecular tilt from the layer normal creates a 2D vector
field of molecular orientation in each layer plane. This field
can be probed optically even in few layer thick films,
enabling the direct observation of orientational spin waves \cite{Young1978}, quasi-long range ordering \cite{vanWinkle1984}, compression induced phase transitions \cite{Link1997,Trittel2017}, topological defects \cite{Pindak1980,Stannarius2016} and defect Brownian motion \cite{Muzny1992}, all in 2D.
Smectics enabled the first electron microscopy of fluid films \cite{Cheng1987}, and x-ray scattering studies of in-plane molecular ordering in smectic films have provided the most convincing evidence to date for the hexatic phase and its melting behavior \cite{Pindak1981}, a scenario emerging from the theoretical work by Kosterlitz and Thouless \cite{Kosterlitz1973} for which they won the 2016 Nobel prize in physics. Stable, thin smectic films can be prepared not only in air, but also in aqueous environment \cite{Iwashita2010,Harth2013}. Many of these smectic film research themes have been reviewed by Pieranski et al.~\cite{Pieranski1993}, and Oswald and Pieranski \cite{Oswald2005}.
The OASIS project focuses on smectic films as a
broad context for the study of a variety of fluid dynamic
phenomena in reduced dimensionality, including 2D
hydrodynamics, interfacial-gradient and applied-field driven
transport in 2D, and the structure and dynamics of
1D interfaces in 2D systems. For example, there are significant
energy barriers to locally changing the film thickness,
since this requires the formation of dislocations associated
with the creation or removal of smectic layers. Thus, any flow in a uniformly thin smectic film is essentially confined to the layer plane \cite{Oswald2005,Picano2000} and conserves the film thickness and density. The continuity equation for the flow field can consequently be reduced to 2D. As a result, a smectic film in a vacuum provides a nearly ideal physical realization of  a 2D incompressible Newtonian fluid \cite{Qi2016}, quite distinct from soap films which exhibit variable thicknesses  \cite{Simons1997}.
If, as is often the case, a smectic film is surrounded by air or some other fluid, motion in the film plane couples strongly to the environment.  Due to the momentum exchange between the film and the embedding fluid, the flow in the film displays features of both 3D and 2D hydrodynamics, behavior that has been dubbed "quasi-2D" hydrodynamics \cite{Oppenheimer2009}.
Thus, a smectic film surrounded by a 3D viscous medium, and containing inclusions such as islands (regions of larger layer number on a few layer thick background film) or particles, is a direct physical analog of protein-containing lipid bilayer membranes.

The hydrodynamics  of inclusions in such thin visco-elastic membranes represents a fundamental physical problem and has been the subject of many experiments in different biological, chemical and physical systems, including smectic films in the context of OASIS \cite{SaffmanDelbruck1975,Saffman1976,Levine_rodmobility2004,Petrov2008,Nguyen2010,Eremin2011,Qi2014}.
More broadly, investigations of the motion of inclusions in smectic films offer the opportunity to explore hydrodynamics in restricted geometries \cite{Nguyen2010,Eremin2011,Qi2014,Qi2016,Kuriabova2016}.  Flow under the influence of external electric and magnetic fields \cite{Cladis1985,CladisSaarloos1992,Cladis1995,Chevallard1999,Chevallard1999b,Link2000,Stannarius2006}, mechanical torques \cite{Mutabazi1992}, and thermal gradients \cite{Godfrey1996,Birnstock2001} have also been studied.  Thin smectic films have also enabled the investigation of hydrodynamic instabilities
\cite{Morris1990,Morris1991,Langer1998}, particle aggregation \cite{Conradi2006,Meienberg2015,Bohley2008}  and quasi-2D emulsions \cite{Bohley2008,Cluzeau2001,Cluzeau2002a,Cluzeau2002b,Cluzeau2003,Voeltz2004,Voeltz2005}.

\subsection{Microgravity experiments}

The vast majority of experiments on freely suspended
smectic films have historically been performed on films
with planar geometry. Gravity-driven motion can be
actively exploited in tilted films to perform microrheology
in 2D \cite{Eremin2011}, but under isothermal conditions,
gravitational effects on horizontal, flat films can be
neglected. If the film geometry is non-planar, however,
any inclusions in the film, particularly in the fluid smectic
A and C phases, will be susceptible to sedimentation under the influence of gravitation. In order to study the dynamics
of such inclusions in non-planar films such as bubbles, one
needs to eliminate gravitational effects. Another reason to
avoid gravity is to simplify the study of films under nonisothermal
conditions. In the presence of gravity, temperature
gradients will necessarily lead to buoyancy-driven air
convection. Since smectic films are so thin, air flow will
readily advect the film material \cite{Birnstock2001}, masking intrinsic thermomechanical effects in the films. Simple evacuation of the film chamber does not solve this problem, because in that case the film establishes thermal equilibrium with the environment by radiation losses, irrespective of the local thermal boundary conditions imposed by the experimenter \cite{Godfrey1996}.

As we shall see below, the influence of the meniscus at
the film boundaries on the dynamics of inclusions is greatly
reduced by creating spherical films supported only by a
thin capillary. Since sedimentation is always an issue in terrestrial
experiments performed in this geometry, this motivates
the need for investigating smectic bubbles under
microgravity conditions.

Microgravity ($\mu$g) can be realized in several ways, the
simplest being drop experiments in a terrestrial laboratory.
Drop towers, parabolic flights, sounding rockets and satellites,
and earth-orbiting platforms such as the International
Space Station (ISS) offer opportunities for more extended
access to reduced gravity conditions, on differing time
scales and with differing degrees of ‘‘weightlessness.” The
use of any of these platforms is contingent on many specific
restrictions. For example, spatial constraints, strong initial
accelerations, the duration of the reduced gravity phase,
and the level of residual accelerations (g-jitter) all affect
the feasibility and design of individual experiments. In
the following, we will describe the realization of 2D fluids
using smectic liquid crystals to perform hydrodynamic
experiments in lg and discuss the technical requirements
and design of these experiments. In particular, we are interested
in the long-term behavior of inclusions in quasi-2D
fluids. The formation, dynamics and coarsening of inclusions
in the form of either islands of excess smectic layers
or isotropic droplets of the mesogenic material embedded
in the liquid crystal film have been studied. We have also
observed the response of smectic bubbles to applied electric
fields, thermal gradients, and mechanical stresses.

\section{Motivation, materials, and microgravity experiments}

Free-standing films a few nanometers thick in the smectic~A (SmA) and smectic~C (SmC) phases are ideal objects for studying hydrodynamic flow patterns and the behavior of inclusions. In these fluid LC phases, the flow is essentially restricted to the plane of the film.
In the SmA phase, the preferential direction of the long axes  of the rod-like molecules (the director, {\bf n}) is along the smectic layer normal, while in the SmC phase, {\bf n} is tilted from the layer normal by an angle $\theta$.
Thin SmA films thus are analogues of $2$D isotropic fluids, where all directions
in the film plane are equivalent, while thin SmC films represent anisotropic $2$D  fluids with
polar nematic order.
%Therefore, such smectic fluids are used in many experiments to study the interaction of inclusions in thin
%membranes \cite{Nguyen10,Eremin11,Cicuta07,Voeltz04,Bohley08,Harth} or to analyze the properties of these liquid
%crystals \cite{Doelle,Qi,Bahr,Eremin2}.
%%In many cases the results of liquid crystal experiments than are compared with other biological or chemical systems \cite{Cicuta07} to analyze more complex problems.\\
Free-standing films of rod-shaped mesogens can readily be prepared with thicknesses from many
 micrometers down to two molecular layers ($\sim 5$~nm). The overall shape of the films depends upon the geometry
of the lateral support and on the
pressure difference (Laplace pressure) between the two film surfaces, and smectic membranes have been studied in planar, cylindrical, annular, and spherical geometries. Freely floating smectic films (not attached to any support) have been used to study the shape-dynamics of closed membranes \cite{May2012,May2014} but
so far only short-term observations ($t \lesssim 1$~s) of such structures have been made.

In most experiments, freely suspended films are bound by a meniscus connecting the film with the support. The shape and extent of the meniscus depend, in general, on the geometry, the material, and the conditions under which the film was drawn, but in most cases the meniscus represents a substantial reservoir of material, with a volume far greater than that of the film itself.
The meniscus can exchange liquid crystal material with the film in order to compensate, for example, for film thickness changes,
growth or shrinkage of islands, changes of the film area, and droplet formation in the films. This exchange is a well-established phenomenon \cite{Oswald2002} that has considerable impact on the film dynamics.
The thickness of the roughly wedge-shaped meniscus typically reaches dozens of micrometers, so its structure is truly three-dimensional.
The meniscus complicates the execution and interpretation of many hydrodynamic experiments. For example, thermally generated convection in smectic films in vacuum was shown to be driven exclusively by thermal gradients in the meniscus \cite{Godfrey1996}, masking intrinsic thermocapillary effects
in the films.
The meniscus influences the motion and evolution of inclusions in the thin films, affecting the structure and evolution of islands on the film by exchanging material with them and thus interfering with Ostwald
ripening of $2$D emulsions formed by islands. There have been many attempts at reducing the effects of the meniscus. One approach is to minimize the length of the film boundaries relative to their diameter by inflating the films to form spherical bubbles \cite{Stannarius1998}. In this case, a meniscus forms only at the opening of the supporting capillary, which can be made quite small (with a circumference on the order of a few millimeters), while the bubble diameter can be in the centimeter range. Since on Earth gravity pulls any objects on a film with non-planar geometry downwards, only microgravity conditions allow the long-term study of the evolution or motion of particles,
liquid droplets and islands on smectic bubbles. In general, microgravitation offers two important advantages for performing hydrodynamic experiments: (1) There is no sedimentation of material, enabling long-term observations
of the motion of inclusions on such curved surfaces. (2) Experiments using thermal gradients are not complicated by the effects of (air) buoyancy, allowing one to observe intrinsic thermocapillary effects.

The aim of the OASIS project ({\em Observation and Analysis of
Smectic Islands in Space}) was designing, preparing, and
conducting experiments on smectic bubbles in the International
Space Station (ISS). OASIS represents the first, and
so far only, set of experiments devoted to the study of thermotropic
smectic liquid crystals in space. In the following,
we will introduce the physical questions relevant to thin liquid
crystal film dynamics that are addressed by microgravity
experiments. We will discuss the specific requirements
and technical conditions for carrying out quasi-2D hydrodynamic
experiments on smectic bubbles and planar freely
suspended films on different microgravity platforms, and
summarize the parabolic flight and sounding rocket experiments
that were conducted in preparation for the ISS
flight.

The main focus of this paper is on describing the OASIS setup and some of the technical issues encountered in the design and execution of the experiments, and showing selected preliminary scientific results of the long-term
studies of island and droplet dynamics carried out on the ISS between June~$2015$ and March~$2016$.
A detailed, quantitative analysis of the flight data will be published elsewhere.

\section{Realization of microgravity conditions for experiments on smectic films}

Experiments on smectic films performed under microgravity can be grouped into three categories:
The first category includes short-time studies of dynamic processes occurring in the sub-second to second range, such as  shape transformations of closed, freely floating smectic membranes. These phenomena were investigated on parabolic flights, which offer up to $22$ seconds of $\mu$g.
The second category of experiments includes studies of thermocapillary effects that can be carried out during the
few minutes of a sub-orbital rocket flight.
The TEXUS rocket experiments facilitated by the DLR ({\em Deutsches Zentrum f\"ur Luft- und Raumfahrt})
provided a flight opportunity with six minutes of microgravity that is described below.
The third category includes long-term observations of inclusion dynamics and the coarsening of two-dimensional island
emulsions. For such experiments, time scales of minutes to hours are needed and the only practical way of carrying out these investigations is on the International Space Station.

Each of these experimental platforms provides different conditions and benchmarks. The main experimentally relevant distinctions
are the duration and quality of the microgravity phase, the constraints on the technical implementation,  repeatability, and the opportunity for direct interaction with the experiment during the \mug phase.
Of course, another criterion is the expenditure of time and resources, with the
ISS experiments requiring a far greater investment of both of these than either the TEXUS mission or the parabolic flights.

A \mug environment is also available in the drop tower at ZARM (Bremen), which provides
9.3 seconds of microgravity of excellent quality. We did not consider this as an option for our experiments
for several reasons. The  acceleration during the initial catapult shot would definitely destroy a thin smectic film.
Performing a simple drop experiment instead would circumvent this problem but this would reduce the
microgravity phase to less than $4.6$ seconds, with at least two hours of preparation required for each drop.
The only experiments where such short microgravity periods could be useful are on the shape dynamics of closed bubbles
but for these experiments the principal advantage of the drop tower, the excellent microgravity quality, is of little relevance.
Parabolic flights, on the other hand, which offer the opportunity of performing $30$ experiments per flight with direct interaction of the
experimenter, were found to be an efficient and productive way of testing equipment and techniques to be used on the ISS.

\subsection{Parabolic Flights}

Parabolic flights were performed on an Airbus A300 at {\em Novespace} in Bordeaux, France. These
flights provided the opportunity to perform many repetitions of the experiment ($30$ per flight) while
allowing the experimenter to continuously observe and control the system.

\begin{figure}[htbp]
 \centering
 \includegraphics[width= 0.7\columnwidth]{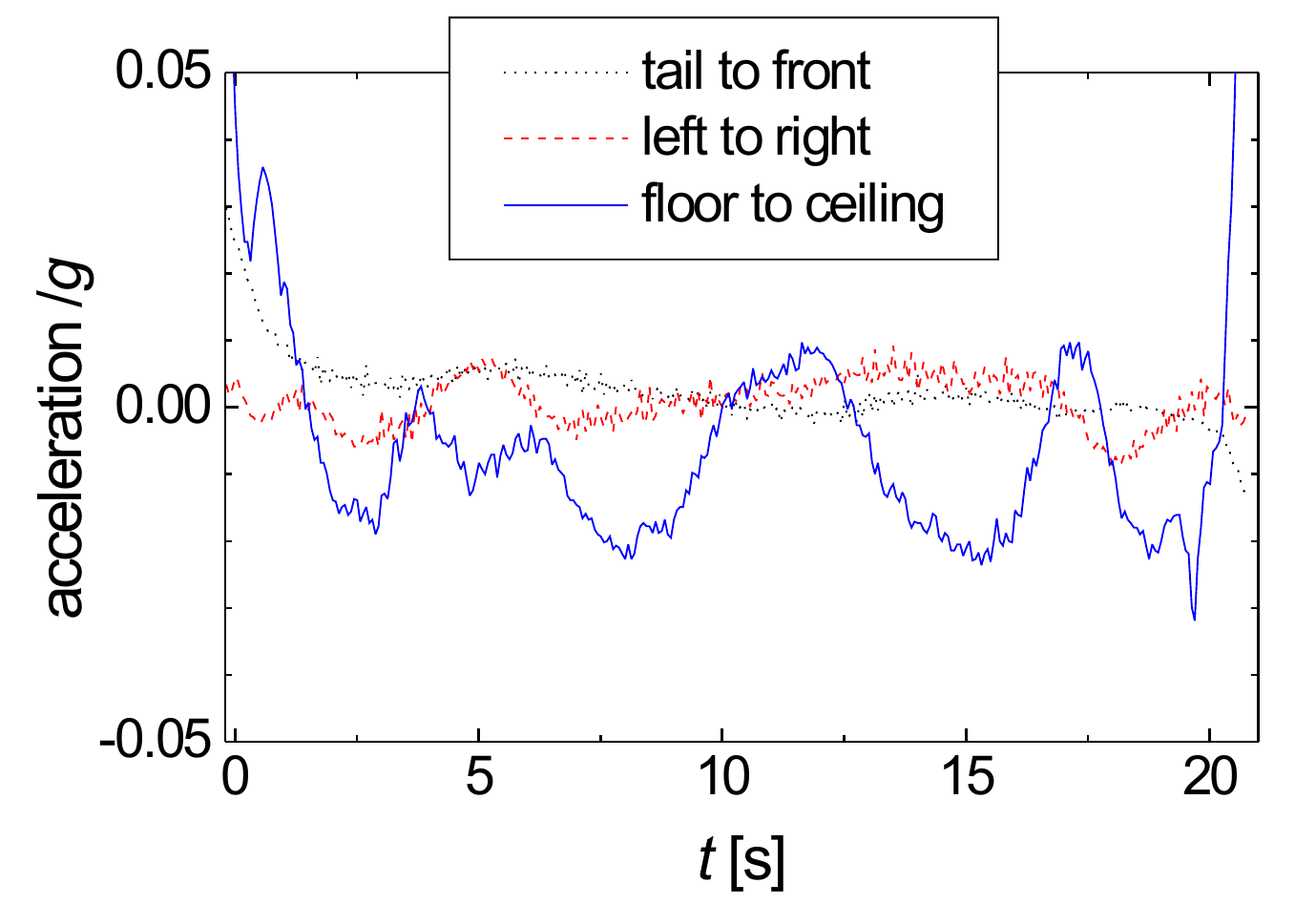}
\includegraphics[width= \columnwidth]{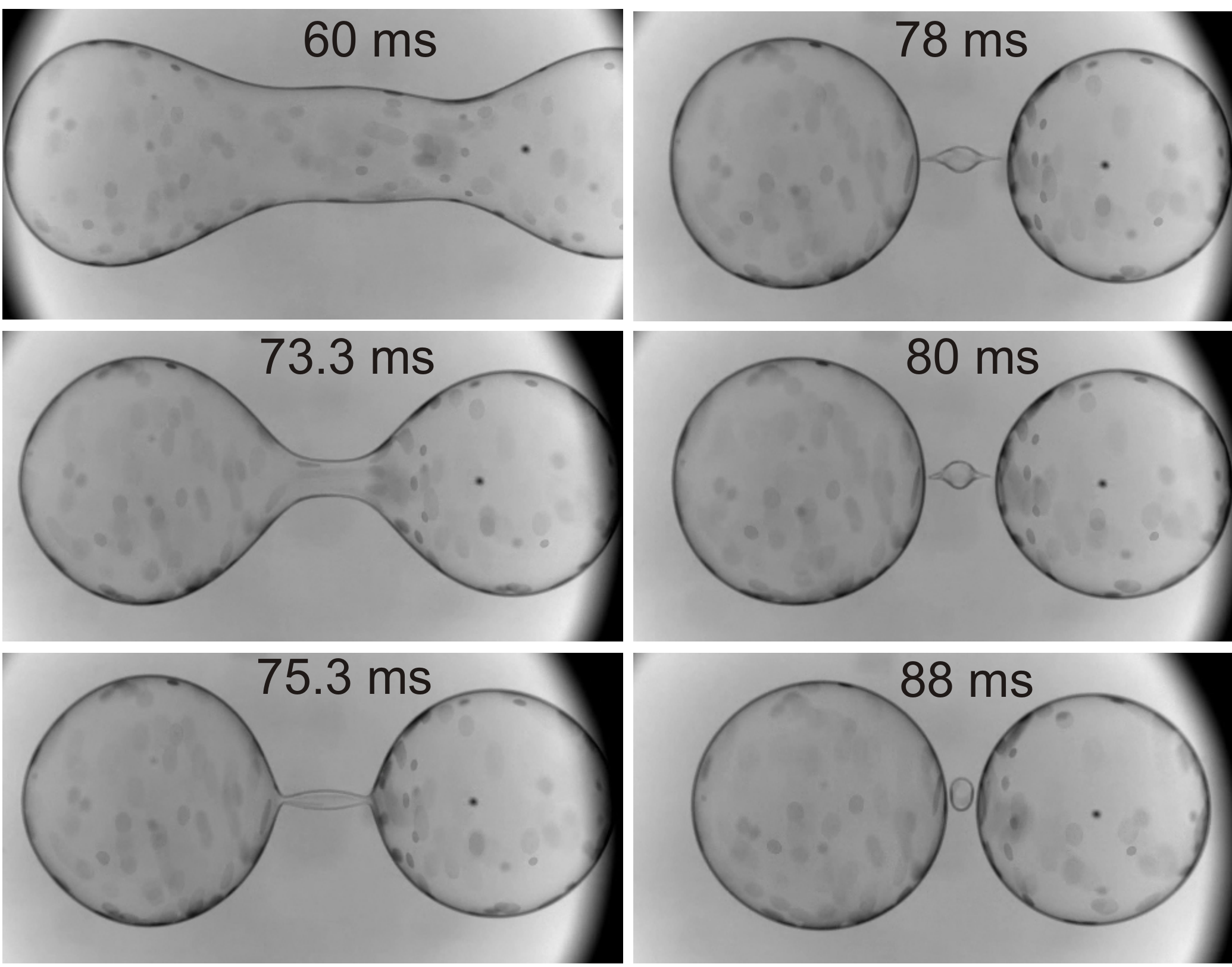}
 \caption{
 Top: Typical acceleration profiles during a parabolic flight (parabola 00 of {\em Novespace} flight 548, on February 11, 2014). The vertical acceleration trace (blue) shows the relatively large deviations from weightlessness that are most problematic for our experiments. The forward and left/right accelerations (green and red curves, respectively) are less critical.
 Bottom: Evolution of a freely floating smectic bubble. The time is given relative to the initial pinch-off of
 the bubble from a collapsing catenoid. A Rayleigh-Plateau instability causes the initially elongated bubble to collapse, becoming dumbbell-shaped before pinching off in the middle, leading to the formation of two large bubbles and a smaller daughter. The image dimensions are $22$~mm $\times 11$~mm. The dark spots are
  islands of excess smectic material floating on the film.
  \label{fig:1}}
 \end{figure}

The $22$ seconds of \mug during each parabola were sufficient to prepare freely floating bubbles from collapsing
catenoid films \cite{Mueller2006,May2012}, and to
observe the bubble shape dynamics or bubble rupture \cite{May2012,May2014}.
The quality of \mug achieved during these parabolic flights is not sufficient
to perform meaningful hydrodynamic experiments with films attached to a solid support, since the $g$-jitter (Fig.~\ref{fig:1}, top) of the airplane, which is of the order of several percent of $g$, directly affects the films.
Freely floating objects, however, are not as susceptible to such disturbances.

The parabolic flights, undertaken primarily during the $20$th parabolic flight campaign of the DLR, were used to test elements of the experimental apparatus to be used in the ISS mission. In addition, we refined the techniques used to prepare free floating smectic bubbles. Short-term observations of smectic islands on the bubble surface and of mechanical instabilities (bubble rupture) were completed mostly during the $27$th and the $29$th DLR parabolic flight missions \cite{May2012,May2014}.

An example of a freely floating SmC bubble, which was prepared using the collapsing catenoid technique \cite{Mueller2006}, is shown in Fig.~\ref{fig:1}.
The mesogenic material is  a 50\%:50\% by weight mixture of 2-(4-n-hexyloxyphenyl)-5-n-octyl\-pyrimid\-ine
and 5-n-decyl-2-(4-n-octyloxy- phenyl)pyrimidine, which is in the SmC phase at room temperature. A Rayleigh-Plateau instability of the initially tube-like closed membrane leads to constriction about its diameter and eventually to the formation of individual, spherical bubbles. As a consequence of the reduction of the bubble surface area by capillary forces, the excess smectic material forms islands. When viewed in parallel, transmitted light, these islands appear as dark spots on the films (see Fig.~\ref{fig:1}). The long-term study of the dynamics of such smectic islands was one of the main goals of OASIS.

\subsection{Sounding rocket experiments}

\begin{figure}[htbp]
 \centering
 \includegraphics[width=\columnwidth]{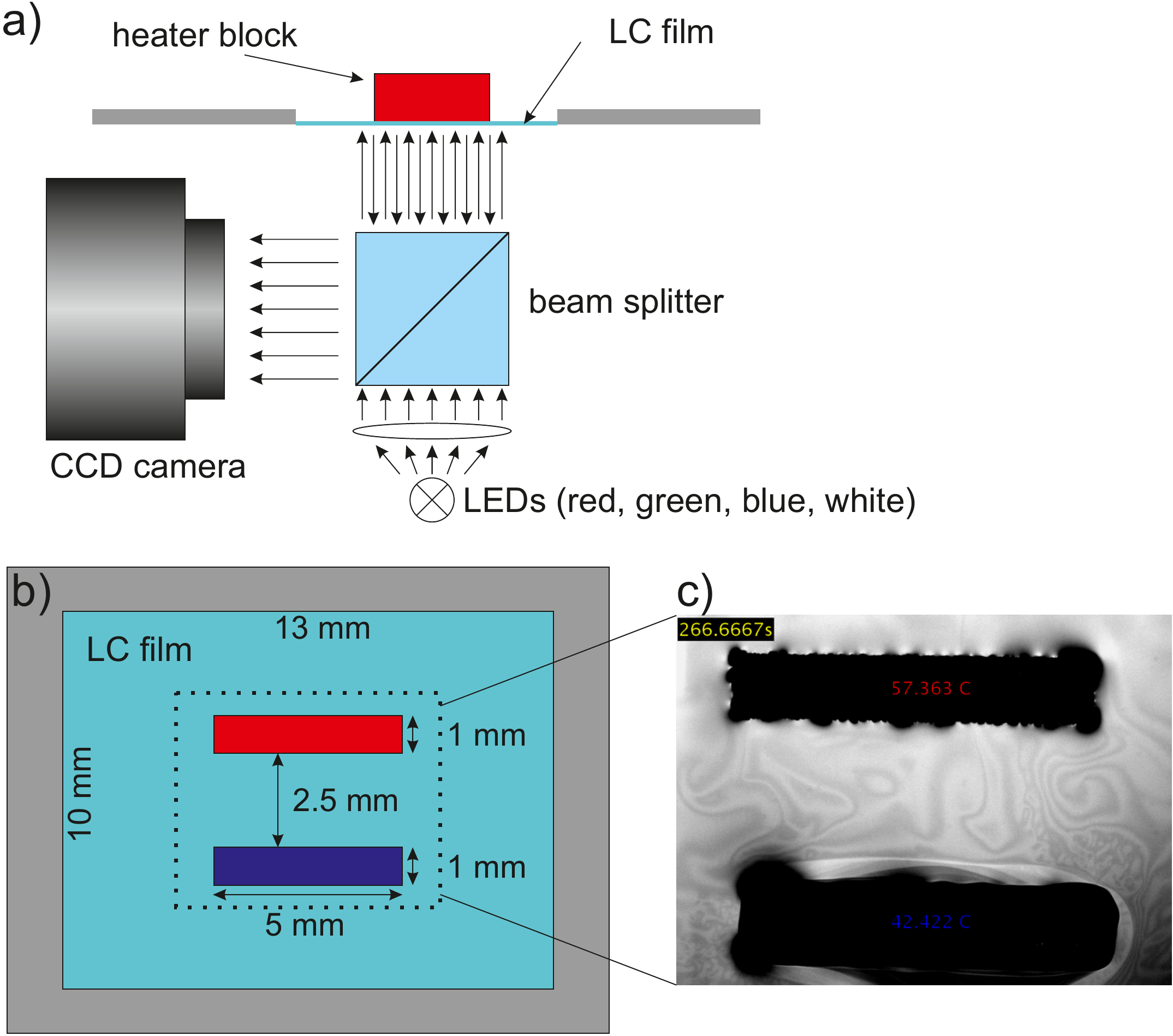}
 \caption{
 OASIS-TEx setup for thermocapillary experiments on smectic films in microgravity. (a) Side view of the illumination system. The film is observed in reflection using monochromatic or white polarized light provide by LEDs. (b) Film and heater geometry. The two rectangular objects (red/blue) are heating/cooling blocks that contact the film. These can be set to temperatures $T_0\pm\Delta T$ above and below ambient to produce temperature gradients of up to $10 \, ^\circ$C/mm in the film. c) Reflection image of Schlieren textures in a 535~nm thick SmC film. Changes in the texture help to visualize thermally-induced  vortex flow, which sets in at the highest temperature gradients achieved in the experiment. The film chamber is at $T_0=50^\circ$C.  Material parameters and details are given in Ref.~\cite{Stannarius2015}.(For interpretation of the references to color in this figure legend, the reader is referred to the web
version of this article.) \label{fig:2}}.
 \end{figure}

With sounding rockets, experimental microgravity times of several minutes' duration can be achieved. This is sufficient to analyze, for example, thermally driven convection in thin, planar liquid crystal films.
As intimated above, smectic films would not withstand the huge acceleration experienced during the launch phase and are therefor only prepared during the flight. The weightless phase is not long enough to create smectic bubbles, which require several minutes of careful
inflation, but planar films can be drawn within less than a minute, leaving enough time to perform
thermocapillary experiments. In preparation for the OASIS mission, the OASIS-TEx (Thermocapillary EXperiment)  was successfully launched with TEXUS~$52$  in Esrange, Sweden on April $27$, $2015$. The rocket reached an altitude of about $250$~km and provided approximately $360$~s of weightlessness of much better quality than available on the parabolic flights.
The TEXUS mission, however, offered only limited opportunities for providing feedback and controlling the experimental parameters during the flight.
These experiments provided the first observations of thermocapillary drift and convection in smectic films
exposed to in-plane temperature gradients in \mug \cite{Stannarius2015}.
The setup for the thermocapillary experiment is shown in Fig.~\ref{fig:2}. LEDs were used for
illumination and the film was observed in reflection using a CCD camera.
Two heating/cooling blocks in contact with the film were used to generate thermal gradients and
changes in the SmC Schlieren texture were used to track flow in the film plane.
The material used in these experiments was $5$-n-octyl-$2$-($4$-n-octyloxyphenyl)pyrimidine, with the mesophase sequence:
Isotropic 68$^\circ$C Nematic 62$^\circ$C SmA 55.5$^\circ$C SmC 28.5$^\circ$C Cryst.

\subsection{ISS experiments}

For long-term experiments on spherical smectic films (bubbles) supported by a thin needle under microgravity conditions, the OASIS setup was successfully launched at the Kennedy Space Center on a SpaceX Falcon 9 Rocket on April $17$, $2015$, and was transported to the ISS in the Dragon CRS-6 module. Experiments were performed until March $2016$. The following sections describe the essential technical details of OASIS and give some preliminary results.

\section{OASIS setup and technical implementation}
\label{sec:setup}

The primary goal of the OASIS experiments on the ISS was to carry out investigations of hydrodynamic flow
in smectic freely suspended films, the relaxation of hydrodynamic perturbations, and the long-term behavior of emulsions of islands and droplets. Experiments were also performed to study thermocapillary phenomena and the effects of applied electric fields and mechanical stresses on the bubbles.
The microgravity conditions on the ISS prevented sedimentation and eliminated buoyancy-driven thermal convection of the surrounding air and in the liquid crystal films.
A bubble diameters of $15$~mm was considered sufficiently large to minimize the influence of
the meniscus around the bubble-inflation needle on the dynamics of inclusions in the film. Film thicknesses were expected to be in the range between about $5$ and several
hundred nm (i.e., from $2$ to $100$ or more smectic layers), with thin films preferred as the background of island emulsions, and thicker films being more suitable
for preparing droplet arrays by melting islands into the nematic or isotropic phase.

We proposed to study emulsions of islands created primarily by shear flow of the films.
Islands can be used as tracers to visualize flow on the bubble surface, and their size and shape can be monitored in order to characterize the dynamical behavior of $2$D island emulsions.
The mutual interactions of inclusions and their interactions with topological and layering defects
are also of interest.
We also planned to shoot picoliter-sized droplets of an immiscible isotropic liquid onto the bubbles using a microdispenser and to study their behavior. In addition to mechanical manipulations of the smectic bubbles by controlled inflation, deflation and shear flow in the film plane, the experimental apparatus was designed to allow the study of the influence of thermal gradients and electric fields.

\begin{figure}[htbp]
 \centering
\includegraphics[width=0.75\columnwidth]{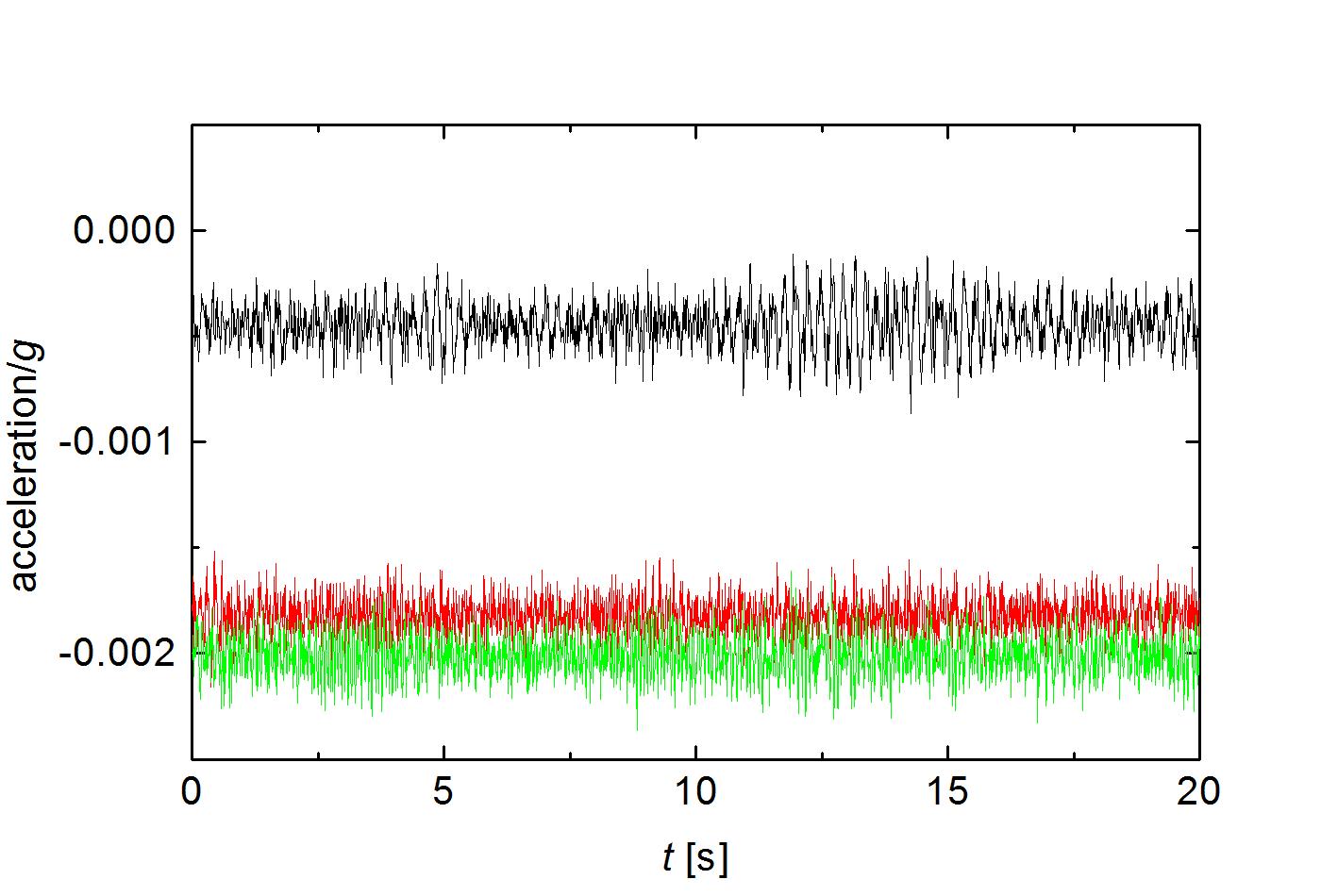}
 \caption{Typical example of acceleration data from the MSG, measured at the OASIS Floor Plate on Nov. 16, 2015,
 starting at 14:35:23.503. Black, red and green  colors (top to bottom curves) refer to the three acceleration components $a_x,a_y, a_z$ in the coordinate system of the sensor. High frequency oscillations ($>50$~Hz) are practically irrelevant for our experiments. The overall quality of \mug is substantially better than in parabolic flights (c.f. Fig.~\ref{fig:1}, top).
 }
 \label{fig:sams}
\end{figure}

Apart from the long-term character of the experiment, the ISS provides an environment where the quality of  microgravity is much better
than in parabolic flights. For comparison with the parabolic flight conditions, we show a $20$~second sequence of typical acceleration data measured
during our experiment in Fig.~\ref{fig:sams}.

\begin{figure}[htbp]
 \centering
\includegraphics[width=0.49\textwidth]{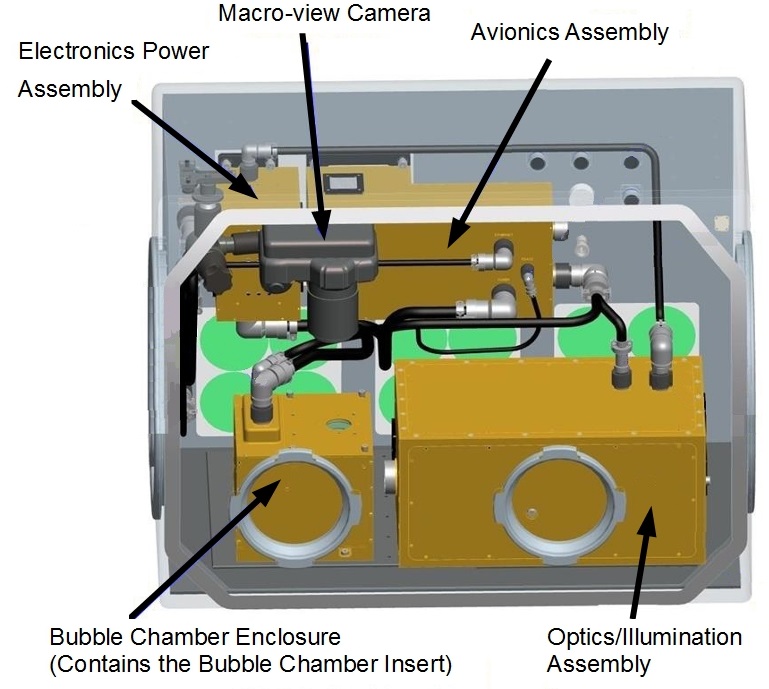}
\includegraphics[width=0.49\textwidth]{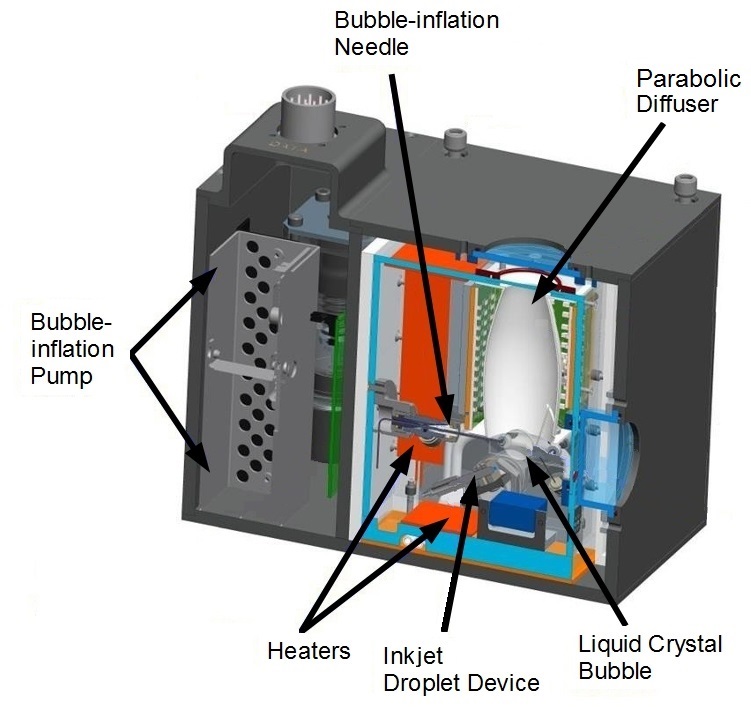}
 \caption{
 The OASIS experiment was housed in the Microgravity Science Glovebox, located in the Destiny Module of the ISS. Left: Overview of the OASIS hardware. Right: Detailed view of the Bubble Chamber enclosure.
 }
 \label{fig:glovebox}
\end{figure}

The OASIS hardware was installed in the Microgravity Science Glovebox (MSG) on the ISS.
A sketch of the MSG with the OASIS hardware installed is shown in Fig.~\ref{fig:glovebox}, left. The main component is the bubble chamber enclosure that contains the experimental setup, including illumination and observation elements, shown at right. The individual elements of the bubble chamber enclosure are described in detail below.

\subsection{Liquid crystal materials}
Four different liquid crystal samples (SN$001$--SN004) were selected for the ISS flight.
A decisive criterion for
selection was that the materials have fluid smectic phases at room temperature and that they could readily be blown into bubbles. We also wished to compare the behavior of
chiral and non-chiral, polar and non-polar, tilted and non-tilted phases.

\begin{itemize}
\item{SN$001$}: Polar SmA (mixture of 8CB + Displaytech MX 12160),\\
 Isotropic  56$^\circ$C  Nematic  54$^\circ$C  SmA  5$^\circ$C Cryst.
\item{SN$002$}: Racemic SmC (Displaytech MX 12846),\\
 Isotropic  84.7--82.0$^\circ$C  Nematic  81.4$^\circ$C  SmA  66.1$^\circ$C SmC
\item{SN$003$}: Chiral SmC$^\ast$ (Displaytech MX 12805),\\
 Isotropic  84.7--82.0$^\circ$C  Cholesteric  81.4$^\circ$C  SmA$^\ast$  66.1$^\circ$C SmC$^\ast$
\item{SN$004$}: Non-Polar SmA (Displaytech MX 12160), \\
 Isotropic   51.1$^\circ$C   SmA  -3.2$^\circ$C Cryst (3.1$^\circ$ SmA).
\end{itemize}

\noindent
$8$CB is $4'$-n-octyl-$4'$-cyanobiphenyl, obtained from {\em Sigma-Aldrich}. The Displaytech materials are proprietary mixtures supplied by the {\em Miyota Development Center of America}.

A separate bubble chamber enclosure and dispenser system was used for each sample. The samples were exchanged periodically by ISS crew members but the experiments were controlled from NASA Glenn Research Center's ISS Payload Operations Center in Cleveland, Ohio. The total amount of each LC sample available for the experiments was about $100~\mu$l.

\subsection{Bubble-inflation needle}
\begin{figure}[htbp]
 \centering
\includegraphics[width=0.4\textwidth]{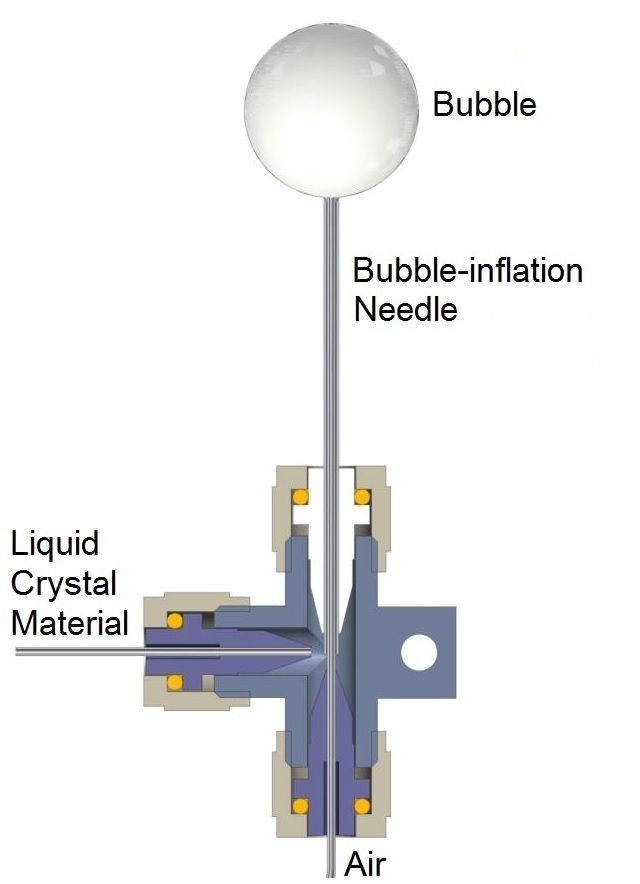}
 \caption{
 Concentric needle device used to generate and inflate liquid crystal bubbles. The inner capillary
  (outer diameter $0.72$~mm, inner diameter $0.41$~mm) is connected to an air pump for inflation and
  deflation of bubbles. The gap between the outer capillary
 (outer diameter $1.27$~mm, inner diameter $0.84$~mm) and the inner capillary is filled with liquid crystal material pumped from a larger reservoir that is used to form an initial thin film over the opening of the inner needle.
 \label{fig:needle}}
 \end{figure}

The primary task in each experiment, and the basis of all subsequent measurements, was smectic bubble generation
using a concentric needle device (Fig~\ref{fig:needle}). At the beginning of each experiment, a small amount of liquid crystal material was driven through the outer needle until a thin cap formed across the
opening of the inner needle. This was done carefully, avoiding excess material that might wet the outer needle surface. If this happened,
the bubble would slide down the needle or become canted in the bubble chamber, in which case the capillary would have to be cleaned by heating the setup far
into the isotropic phase of the mesogen. Once a cap had formed, air was pumped
through the inner needle to generate a thin, oriented film
which was then gently inflated to become a spherical bubble.
Bubble inflation was carried out at high temperatures in the smectic phase, where the viscosity of the
material is low, achieved by heating the needle assembly using a Nichrome wire heater wrapped around it.
During inflation, the needle temperature was set in the range from $25^\circ$C--$75^\circ$C, depending on the liquid crystal material.
The bubbles were inflated using a {\em Cavro Xcalibur} $30$~mm Stroke Syringe Pump.
The maximum pump rate was $0.83$~ml/sec, although
in order to inflate thin, stable bubbles, the flow-rate could be reduced to
$0.21 \, \mu$l/sec. In general, the inflation speed determined the film thickness, with faster inflation leading to thinning of the films.
Rapid deflation of the bubbles (for example, in order to create islands) was limited by the pumping rate. With
bubble volumes of almost $2$~ml, the fastest possible deflation took a few seconds.

\subsection{Macro-view and micro-view cameras}

\begin{figure}[htbp]
 \centering
\includegraphics[width=0.7\textwidth]{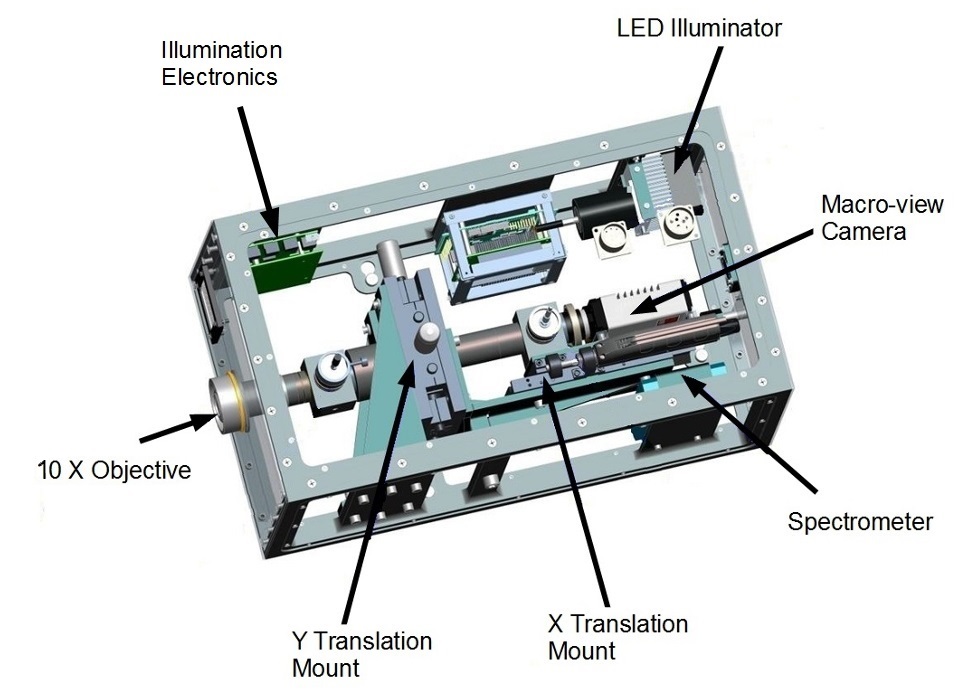}
 \caption{Optics/illumination assembly with micro-view camera, spectrometer, objective and LED illuminator. \newline \label{fig:opt}}
 \end{figure}

All visual observations of the bubbles were made using two identical, high-resolution, color CCD cameras ({\em Prosilica GX 1050C, Allied Vision Technologies}). An external optics/illumination assembly (shown in Fig.~\ref{fig:opt}) housed the micro-view camera and a spectrometer used for film thickness measurements. The micro-view camera  was mounted on a simple reflected light microscope equipped with a $10\times$ objective ({\em 10x Mitutoyo Plan Apo SL Infinity Corrected Objective,  Edmund Optics PN NT46-144}) used to obtain detailed images of film textures. The objective, which has a $15$~mm working distance, was positioned to image the top of the bubble located between the airjets, with a field of view of $500 \times 500 \, \mu{\mathrm m}^2$ and a depth of field of $3.5~\mu$m. The camera could be translated laterally using two linear positioners. The cameras support a maximum frame rate of $112$~fps but in the OASIS experiments, a maximum rate of $60$~fps was used. The micro-view camera allowed us to observe such phenomena as the coalescence and coarsening of islands, and hydrodynamic effects caused by applied electric fields, thermal gradients and the airjets, with high resolution.
The numerical aperture of the micro-view camera was $0.28$ and the spatial resolution about $0.5~\mu$m  ($1024\times 1024$ pixels).
The spectral sensitivity range of the camera was from $350$~nm to $1000$~nm.

The system could be converted to a polarizing microscope by inserting linear glass polarizing filters ({\em Edmund Optics PN TN43-785}) with an efficiency of $95\%$. The LED illuminator ({\em DiCon G180 Series ScopeLED}) had a color temperature range of $2700$~K to $7500$~K (we selected $6500$K) and a maximum output of $975$~lm. For interferometric measurements of the thickness of the films and islands, a spectrometer ({\em StellarNet BLUE-Wave PN UVN}) with a wavelength sensitivity in the range $250$~nm to $1100$~nm was available. The spectral resolution was $6$~nm with the option to improve it to $1$~nm.

\begin{figure}[htbp]
\centering
 \includegraphics[width=0.7\textwidth]{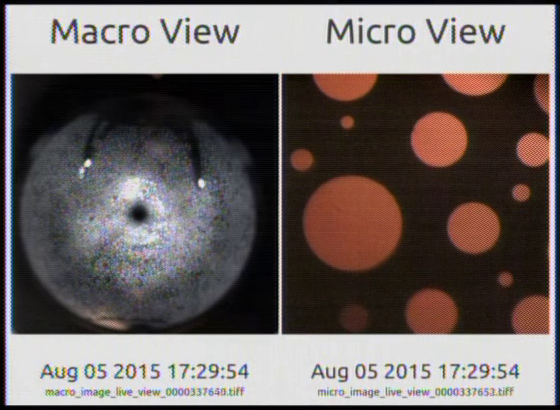}
 \includegraphics[width=0.7\textwidth]{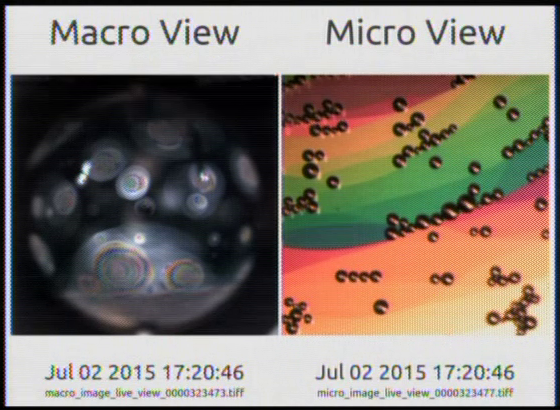}
\caption{
Typical macro- and micro-views of the smectic bubble experiment captured from the live downlink.
The macro-view images show the entire bubble. The bubble-inflation needle is just out of frame at the bottom of these images and curved reflections of the airjet needles can be seen near the top of the upper bubble. The corresponding micro-view images show smectic islands on a thin background film (top) and molten droplets pinned to layer steps in a thick background film (bottom). The different interference colors are indicative of the film thickness.
The resolution of the downlinked images is $512\times 512$~pixels.
\label{fig:downlink}}
 \end{figure}

In addition, each bubble chamber enclosure was equipped with
a macro-view camera (see Fig.~\ref{fig:glovebox}), which imaged the entire bubble. Achieving uniform illumination of the bubble was one of the biggest challenges in designing the OASIS hardware. The bubble was positioned inside a translucent, parabolic diffuser that was illuminated from the outside by several arrays of LEDs. The macro-view camera allowed us to observe the bubble during the inflation stage (and monitor its diameter afterwards), and was used to record large-scale phenomena such as
thermocapillary flow, coarsening dynamics (including Ostwald ripening and the coalescence of smectic islands
and thinner, circular domains in the film called ``holes''), and hydrodynamic instabilities in $2$D.
The macro-view camera was equipped with a $50$~mm f/$2.8$ macro lens ({\em Sigma EX DG Macro, PN 346101}) and provided an image with much lower magnification than the micro-view camera,  the resolution being about $18~\mu$m per pixel.
During the flight, a data downlink provided low-resolution images from both cameras, examples of which are shown in Fig.~\ref{fig:downlink}.

\subsection{Airjet needles}

\begin{figure}[htbp]
 \centering
\includegraphics[width=0.5\textwidth]{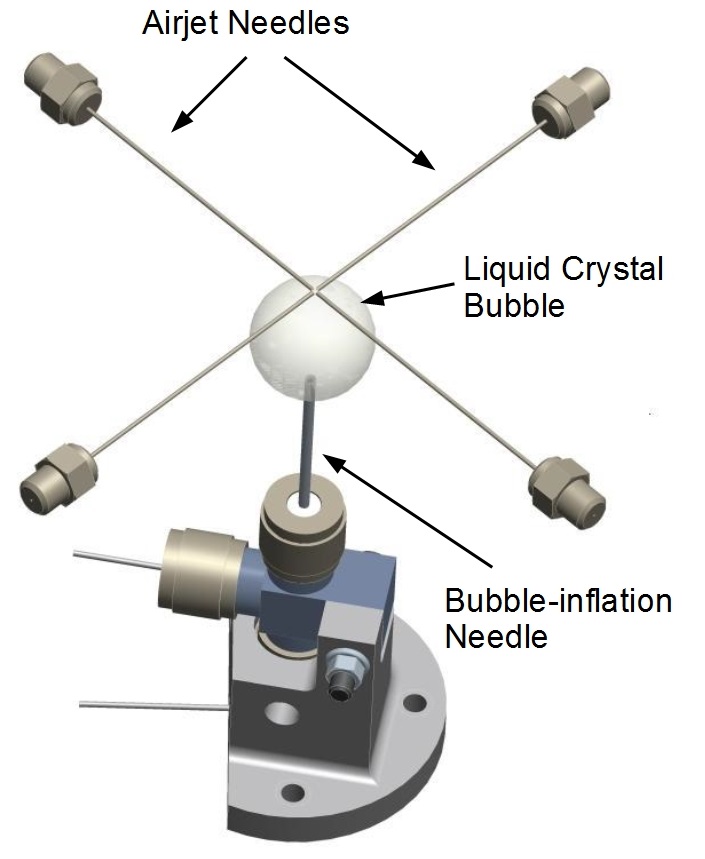}
 \caption{
 Airjet and bubble-inflation needles. Four independent airjets oriented tangentially to the top of the bubble  allowed the creation of flow in arbitrary directions along the bubble surface. One of the needles could be heated to perform
 experiments under non-isothermal conditions. Two of the needles also served as electrodes to apply
 in-plane AC or DC electric fields to the smectic film. The needles were fixed in the chamber \DEL{, just above the top of the bubble,}
 the distance between the airjets and the smectic film depending on the bubble diameter.}
 \label{fig:airjets}
 \end{figure}

Several techniques could be used to create islands on uniformly thick smectic films. The most convenient was the application of strong extensional air flow along the film surface using
four air jets, shown in Fig.~\ref{fig:airjets}. Filtered station air
(supplied by the MSG) blown along the bubble surface led to continuous flow of material all around the bubble. As a result of this flow, excess material in the meniscus could be drawn onto an initially uniform bubble and then broken up into an emulsion of small islands. Alternatively, existing arrangements of islands and holes could be redistributed or randomized in order to reinitialize the system and start a new experiment. Hydrodynamic instabilities could also be triggered. The airjet needles provided controllable air flow of between $10$ and $150$~sccm (standard cubic cm per minute).
In addition, an electric field could be applied between two of the opposing needles, whose tips were about $2$~mm apart. A third needle could be
heated to generate temperature gradients on the bubble surface (further details are given below).

\subsection{Temperature control}

The bubble chamber could be heated uniformly above station temperature (which varied between $20$ and $25^\circ$C)  up to
$60^\circ$C $\pm$0.5$^\circ$C. This was achieved using Kapton (polyimide) encapsulated electrical heaters ({\em Omega KHLV-103 Series}) distributed on the aluminum bubble chamber walls.
%\TODO{Magdeburg ms used to say ``P/N K$05711980$ Series heaters'' but not sure what these are}

Temperature control allowed the study
of the temperature dependence of dynamic parameters under isothermal conditions. We were also able to heat bubbles to the smectic--nematic or smectic--isotropic transition, where layer-by-layer
thinning of the films (melting of the inner layers and redistribution of the molten material) leads to the creation of
nematic or isotropic droplets on the film. The self-organization of those droplets, their dynamics, and
their evolution as a function of temperature, were some of the interesting phenomena that could be
explored under microgravity conditions.

An important goal of the OASIS project was to investigate
the effects of thermal gradients on thin films and any
embedded inclusions. The temperature dependence of the
surface tension of the LC material and changes in the
orientational order parameter were expected to contribute to thermally-induced drift and convection in the films.
Buoyancy-driven convection was eliminated in zero gravity.
Temperature gradients could be applied by heating
either the bubble-inflation needle or one of the airflow needles
to higher temperature than the rest of the chamber.
For this purpose, one airjet needle was wrapped with
Nichrome wire, which allowed us to heat the needle to a
maximum of $75^\circ$C. The distance between the heated airjet
needle and the bubble surface was determined by the bubble
diameter. Temperature differences of about $30^\circ$C could be realized across the $15$~mm bubble.

\subsection{Electric fields}

Liquid crystal materials couple dielectrically to applied electric fields. In chiral, tilted mesophases there is also ferroelectric coupling to the spontaneous polarization. In SmC and SmC$^\ast$ phase films, applied fields cause reorientation of the in-plane tilt direction, an effect that is clearly visible in a reflected polarized light microscope. This reorientation of the director field may lead to the rearrangement of islands or droplets in such films.
Any charged inclusions in either SmA or SmC films will experience Coulomb forces in an applied field, and the presence of ions in the liquid crystal material may lead to electrically driven convection, especially in low frequency or DC fields \cite{Morris1990,Morris1991,Langer1998}.
Two of the airflow needles were therefore configured as electrodes. The tips of the needles were about $2$~mm apart and a maximum voltage of $100$~V (DC) could be applied between them. The airjets were typically located  $\sim 1$~mm above the bubble.
The applied voltage was generated using an operational amplifier circuit (included in the avionics package of the MSG). The frequency could be varied from DC to $10$~kHz, in steps of $10$~Hz between $0$ and $1$~kHz and in steps of $100$~Hz between $1$~kHz and $10$~kHz.

\subsection{Droplet dispenser}

In addition to the observation of smectic islands and holes on the bubble surfaces, we proposed to
investigate the behavior of isotropic liquid droplets embedded in the film.
The plan was to shoot micrometer-sized glycerol/water droplets (from a mixture of water and less than $5\%$ glycerol) onto the bubble surface using an inkjet droplet dispenser \cite{Doelle2014}.
The dispenser ({\em MicroFab Technologies MJ-ATP-01-030-DLC}) had a $30~\mu$m dispensing diameter and was designed to eject
between $1$ and $500$ droplets per second. This dispenser was positioned near the base of the bubble, and could be rotated azimuthally using a stepper motor. The inkjet device
failed to work during the ISS mission, with no droplets generated at all. Instead, we produced
droplets by heating the material to the upper limit of the smectic temperature range, as described above. This approach is not straightforward,
with the heating protocol (both the heating rate and final temperature) being critical, and the bubbles highly susceptible to rupture near the phase transition. Such thermally induced droplets do not nucleate homogeneously, appearing in random positions in the film, and it was not possible to confine them to certain areas or to distribute them homogeneously on the bubble surface. Their size distribution, however, turned out to be rather uniform. This method of creating droplets by heating could only be carried out with the two SmA samples because they had phase transition temperatures within the nominal operating range of the bubble chamber. With the SmC materials, the transition temperatures were too high.

\section{Preliminary results}

The scientific output of the TEXUS and parabolic flight missions performed in preparation for the OASIS mission have been outlined above.
The majority of the OASIS experiments performed on the ISS were very successful and most of the scientific goals were achieved.
The OASIS hardware was installed in the MSG by Cosmonaut Gennady Padalka on June~$21$, $2015$. Sample changeouts, hard drive swaps, and glycerol/water fills were subsequently performed by Cosmonaut Oleg Kononeko and Astronauts Scott Kelly, Tim Peake, and Tim Kopra.
Low-resolution downlink video data have been available since the
start of the mission in March $2015$.
Hard drives containing high definition imagery
recorded on the ISS have recently been returned to Earth and will be used to analyze the experiments in detail.
The following preliminary results are based on an initial, qualitative evaluation of the downlink data.

\begin{figure}[htbp]
\centering
\includegraphics[width=0.9\columnwidth]{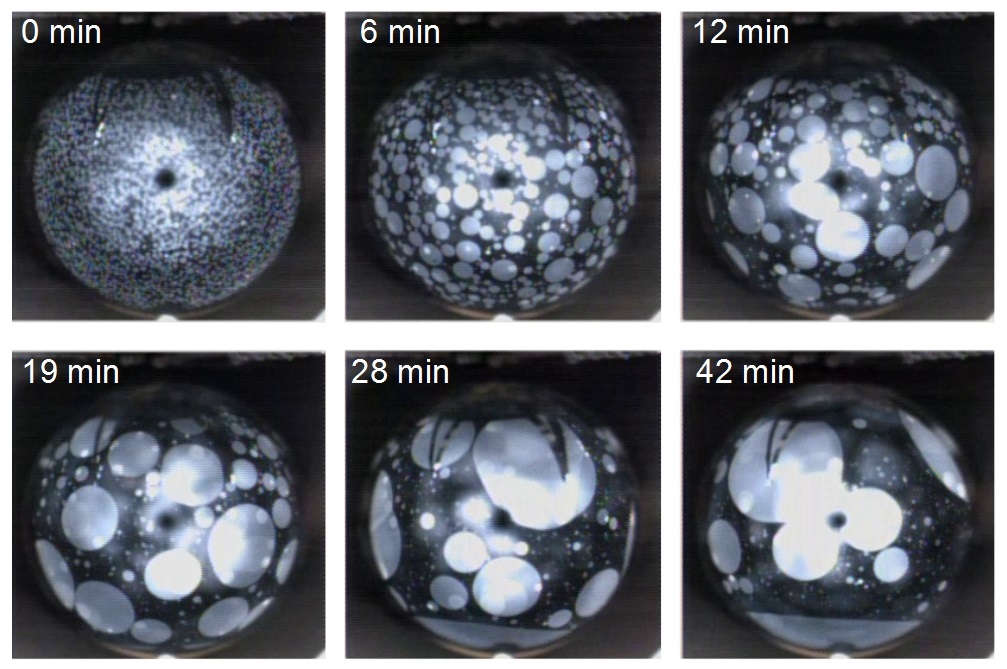}
\caption{
Coarsening dynamics of smectic islands on a thin SmA bubble observed over a period of $42$ minutes. At $t=0$, the initial $2$D island emulsion, prepared by shear flow using two opposed airjet needles, is a fairly monodisperse collection of small, disk-shaped smectic inclusions. The emulsion coarsens over time, by Ostwald ripening and coalescence. The bubble
is approximately $15$~mm in diameter.
The tip of the  bubble-inflation needle is visible at the bottom of these images.
The material is SN001 and the chamber temperature $35^\circ$C.
}\label{fig:coarsening}
\end{figure}

\begin{figure}[htbp]
\centering
 \includegraphics[width=0.8\columnwidth]{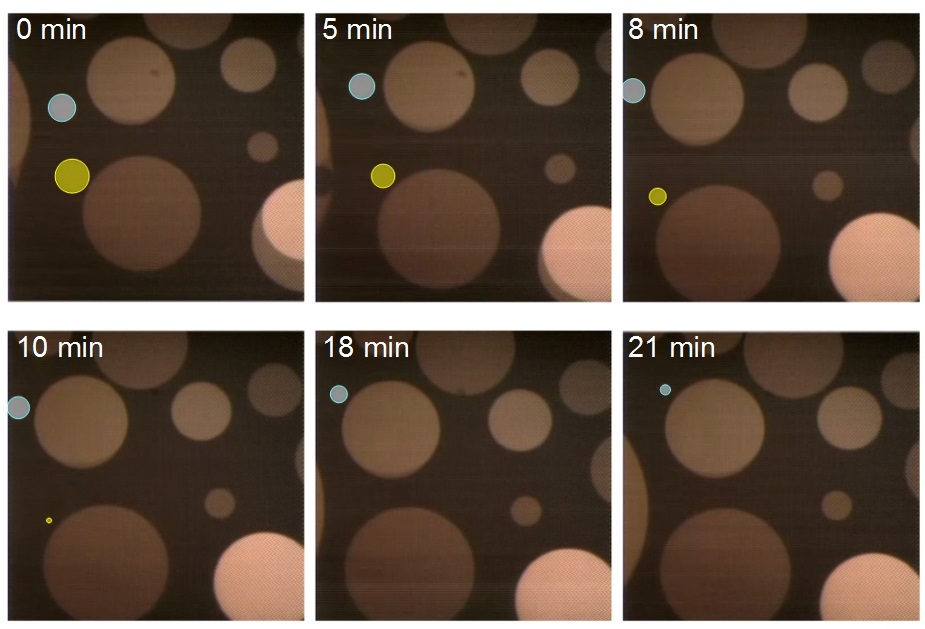}
\caption{
Ostwald ripening of a SmA island emulsion observed with the micro-view camera over a period of $21$~minutes. Two small islands (highlighted here in false color) shrink and disappear.
The material is SN$001$ and the chamber temperature $24^\circ$C.
\label{fig:Ostwald2}
}
\end{figure}

\paragraph{Coarsening of $2$D emulsions of smectic islands}
Macro-view images recorded during a $42$ minute coarsening experiment of smectic islands are shown in Fig.~\ref{fig:coarsening}. The average initial island diameter can be controlled by the amplitude and duration of the shear induced by the airjet needles, with the distribution of island sizes immediately
after the shear flow is stopped being rather narrow. On the ISS, the Perrin length $L=k_B T/(mg)$, a characteristic height associated with sedimentation of particles with mass $m$ under gravity, is much larger than the bubble diameter (on Earth, it is comparable to the island diameter).
The islands are initially distributed fairly uniformly on the bubble surface, as seen in the first image of Fig.~\ref{fig:coarsening}.
The line tension associated with the dislocations at the island boundaries causes the islands to be circular.  The dark background film is only a few smectic layers thick, while the islands have dozens of layers.
Over time, the island emulsion coarsens significantly, with the average island diameter increasing and the total number  of islands decreasing. This is achieved both by Ostwald ripening (the effective diffusion of LC material from smaller islands through
the uniform thin background film to larger islands) and by island coalescence. In the experiment depicted in Fig.~\ref{fig:coarsening}, a large island eventually becomes trapped at the bubble-inflating needle. An example of the redistribution of island material by Ostwald ripening recorded with
the micro-view camera is shown in Fig.~\ref{fig:Ostwald2}. Here we see two small islands located between several larger ones shrinking and disappearing over a $21$~minute period.

\begin{figure}[htbp]
\centering
bottom heating\\
\includegraphics[width= 0.7\columnwidth]{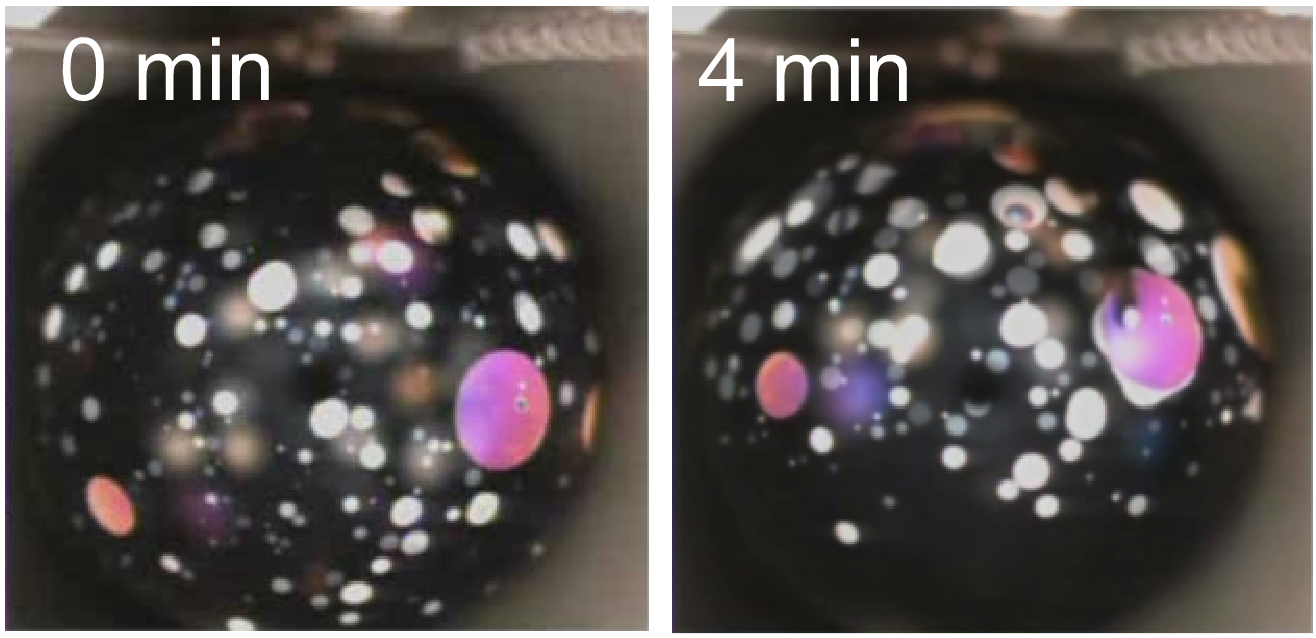}
\includegraphics[width= 0.7\columnwidth]{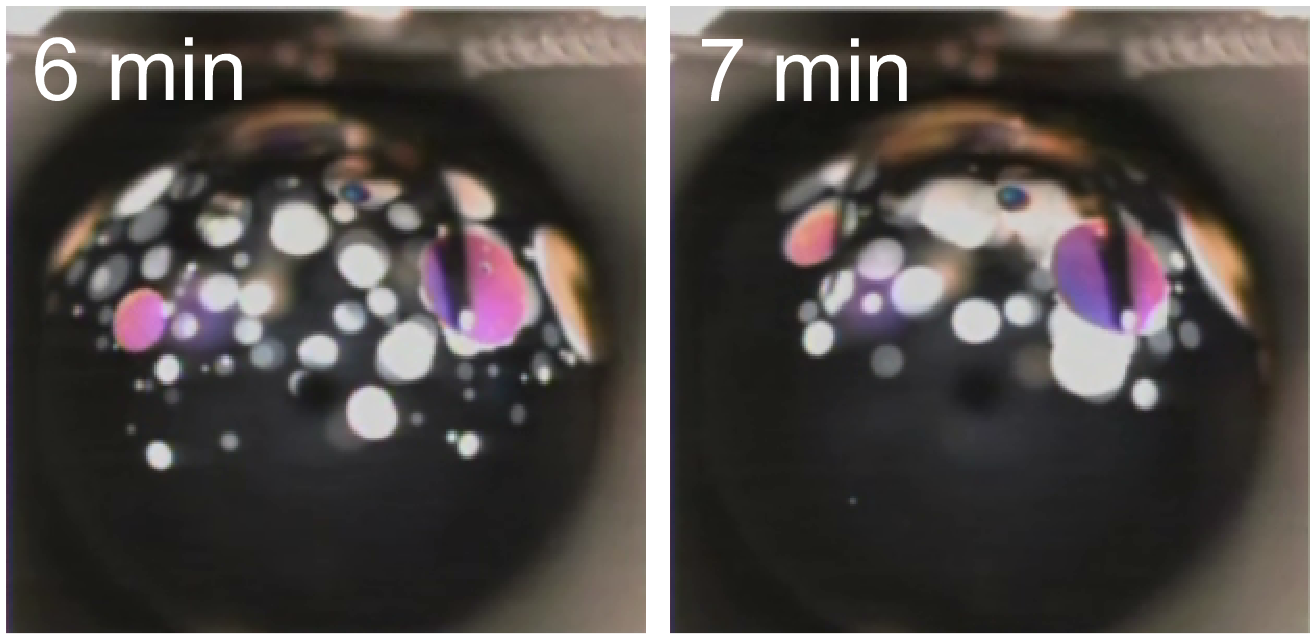}\\

top heating\\
\includegraphics[width= 0.7\columnwidth]{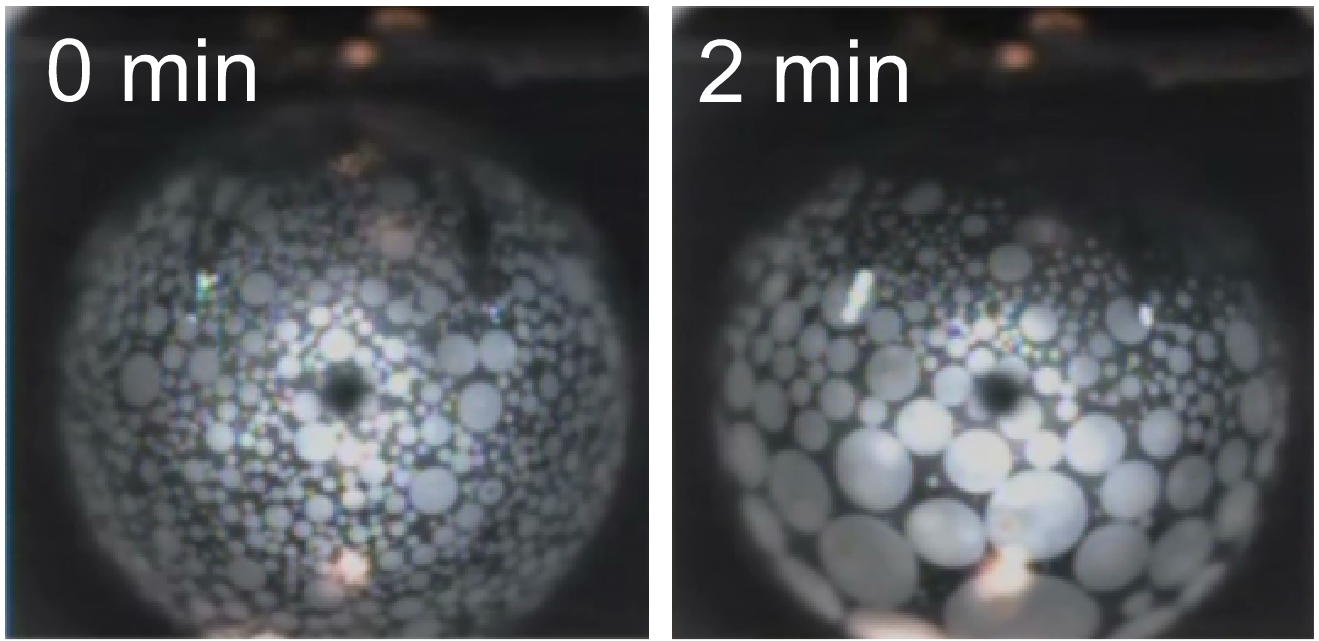}
\includegraphics[width= 0.7\columnwidth]{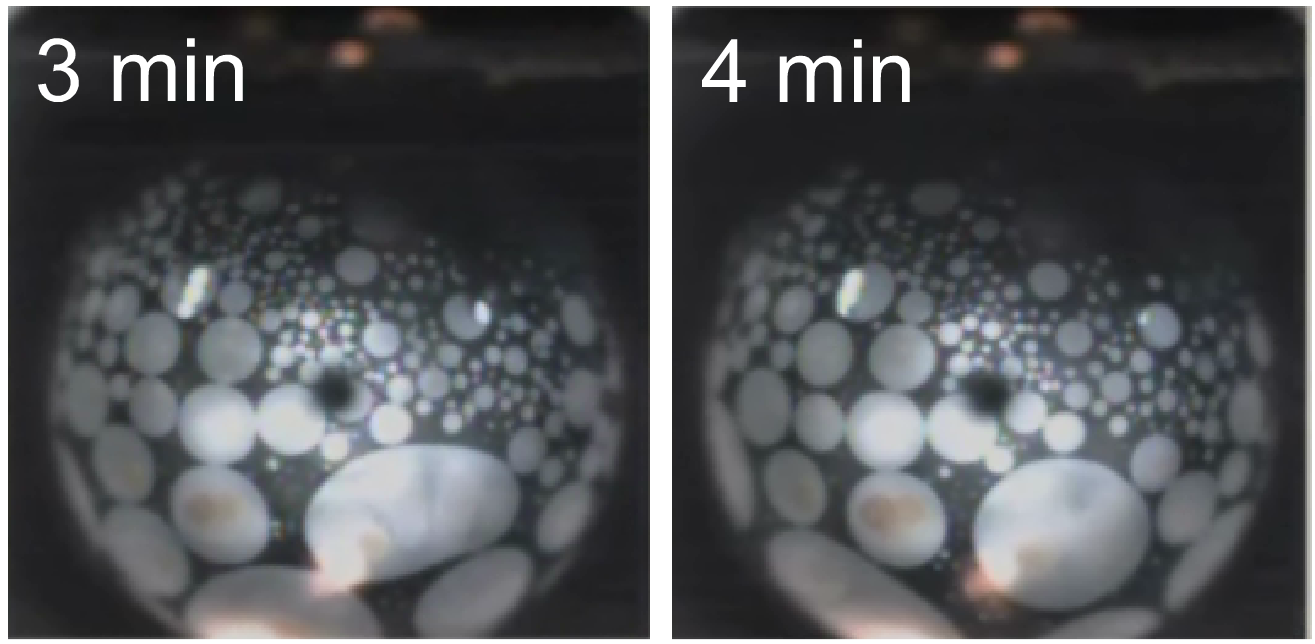}
\caption{
Thermocapillary flow in a SmA bubble caused by heating the bubble-inflation needle, located just out of frame at the bottom of the bubble (top sequence), or by heating two of the airjet needles, located just above the top of the bubble (bottom sequence). In both cases, the islands are transported to the
colder side of the bubble, against the temperature gradient.
The material was SN$004$ and the chamber temperature $25^\circ$C.
The bubble-inflation and airjet needles were heated in turn to $45^\circ$C.
}
\label{fig:thermal}
\end{figure}

\paragraph{Thermally-induced island migration}
When the bubble-inflation needle is set to a temperature higher than the environment temperature of the chamber, a slow, thermally-induced migration of objects on the film against the temperature gradient,  towards the colder regions, is observed. The islands move with velocities on the order of $10 \, \mu$m/s. When, instead, two airjet
needles near the top of the bubble are heated, the direction of migration is reversed, with the islands now moving
towards the bottom of the bubble.
Typical scenarios of island reorganization in a bubble in such temperature gradients are shown in Fig.~\ref{fig:thermal}.

\paragraph{Self-organization of droplets}

\begin{figure}[htbp]
\centering
 \includegraphics[width=0.41\columnwidth]{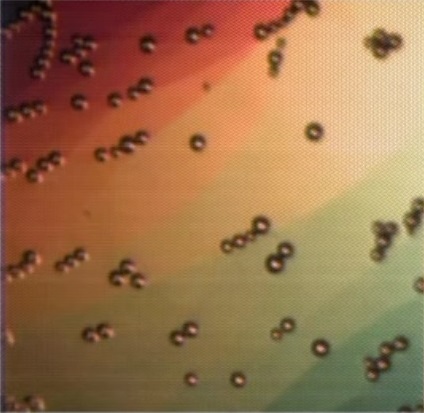}
\includegraphics[width=0.40\columnwidth]{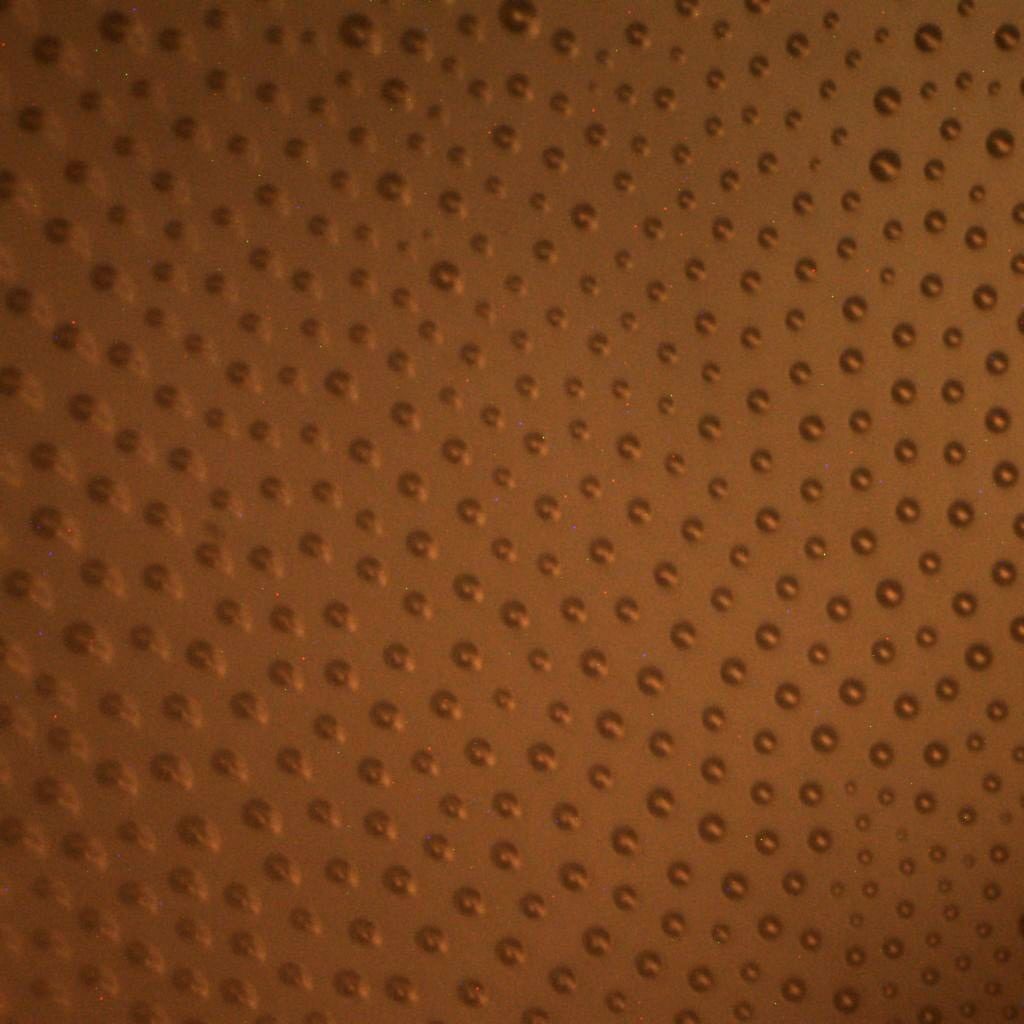}
\caption{
Left: Small nematic microdroplets (with diameters of around $50~\mu$m) assemble
and form chains at layer
steps in an inhomogenously thick SmA film with terraces. This is a consequence of capillary interactions \cite{Schuring2002,Schuring2004}.
The material is SN$001$ and the chamber temperature is $55^\circ$C.
Right: In uniformly thick SmA films without layer steps, molten isotropic microdroplets self-organize in a more or less regular hexagonal lattice.
The material is SN$004$ and the chamber temperature is at the clearing point (smectic--isotropic transition) of the bulk material. In the films, this transition occurs at slightly higher temperatures than in the bulk.
These images were recorded with the micro-view camera.
}
\label{fig:droplets}
\end{figure}

Droplets created in SmA films are subject to two self-organizing processes. In inhomogeneous films, the droplets tend to cluster along layer thickness steps, on the thicker plateau but pinned to the dislocation  \cite{Schuring2002,Schuring2004}, as shown in Fig.~\ref{fig:droplets}, left. In SmA films of uniform thickness, \RV{molten} droplets are seen to self-assemble into remarkably regular, hexagonal lattices, illustrated in Fig.~\ref{fig:droplets}, right.
The nature of the mutual interactions between the droplets leading to this latter behavior is not understood at this time.

Further island/droplet experiments were performed by rapidly shrinking then re-inflating the bubbles.
Rapid shrinkage leads to a quick reduction of the film area that cannot be compensated
by flow into the bubble meniscus. The excess material instead forms islands on the films. One of the aims of the OASIS
experiment was to investigate the influence of deflation speed and deflation ratio on the size distributions
and densities of the resulting island emulsions. These experiments should provide insight into dislocation dynamics in smectics.

\
\section{Summary}

\subsection{Achievements}

Parabolic flights provided the opportunity to perform
experiments on freely floating, closed smectic bubbles and yielded interesting results on
their shape transformations and rupture dynamics \cite{May2012,May2014}. The sub-orbital TEXUS mission allowed us to observe thermocapillary effects in thin free-standing films in air in $\mu$g.
The data from these experiments have been evaluated and the findings are being prepared for publication.

The ISS experiments led to novel insights into the coarsening dynamics of $2$D emulsions of
smectic islands on spherical films, which were seen to proceed both by island coalescence and Ostwald ripening.
Island formation and disappearance during bubble deflation and inflation, respectively, were observed and recorded.
Migration of inclusions in thermal gradients established across the bubble, and the ordered self-assembly
of microdroplets on the curved bubble surface, were observed.
Following the successful return of the OASIS hard drives to Earth, detailed evaluation of the data is now in progress.

\subsection{Technical problems}

Several technical problems had to be overcome during the ISS experiments. The biggest issue was the malfunction of the droplet dispenser. All attempts at shooting glycerol/water droplets onto the bubbles failed. It appears that during the mission, the inkjet device \DEL{probably} did not produce any droplets at all, for
reasons that are not clear. While under normal gravity it was found to be quite straightforward to fill the dispenser with
bubble-free fluid, microgravity conditions on the ISS may have led to incomplete filling, with air trapped in the
dispenser body preventing droplet ejection. A workaround during the mission was the creation of droplets from molten smectic material.
This was possible for only two of the flight samples, however, the phase transition temperatures of the other two liquid crystal materials exceeding the designed temperature range of the bubble chamber.

In general, focusing the micro-view camera turned out to be a tedious and time-consuming procedure, particularly whenever a new bubble was inflated or the bubble diameter changed. Once the focal plane was established, however, the
micro-view camera worked very well.
Towards the end of the mission, the micro-view camera experienced another technical issue, when malfunction of the linear positioners prevented the camera from being translated to different locations, which greatly restricted its usefulness.

Overall, these problems demanded some improvisation and flexibility in carrying out the planned experimental procedures but they hardly affected the successful outcome of the OASIS mission.

\subsection{Outlook}
The OASIS experiments have demonstrated the advantages of microgravity in performing fundamental investigations of thin, smectic liquid crystal films. It is already clear, even before any quantitative evaluations have been completed, that these experiments represent only the beginning of the exploration of this field. Many new questions have arisen,
related to the basic physics of liquid crystals and the hydrodynamics of thin, fluid films. For example, the experiments suggest ways of exploring the local dynamics of
freely floating films, where processes like budding and wrinkling have been observed \cite{May2012,May2014}. Further experiments should help elucidate their dependence on the fundamental physical properties of smectic membranes. Thermocapillary
experiments in \mug have revealed unexpected behavior in planar films. In future experiments, studies under well-defined thermal conditions are needed to understand quantitatively the fluid migration induced by temperature gradients in quasi-$2$D smectic films. A TEXUS reflight
to continue these investigations is scheduled for $2017$. Finally, the study of droplets of immiscible materials on
smectic bubbles is in its infancy. Of particular interest would be to investigate the motion of inclusions on films
with non-uniform Gaussian curvature, such as catenoids or Delauney surfaces in general. In such experiments, gravitation generally masks all capillary effects in uniformly thick films. With a slight modification of the
 OASIS setup, all these studies would become feasible in microgravity.

\section*{Acknowledgments}
The authors acknowledge financial support for the
OASIS project from NASA Grant NNX-13AQ81G, and
from the DLR within projects OASIS-Co 50WM1127
and 50WM1430. We are particularly indebted to the
DLR for making the TEXUS and parabolic flights possible,
and to Airbus DS for the construction and testing of
the OASIS-TEx equipment, as well as for technical support
during the TEXUS $52$ campaign. ZIN Technologies, Inc. is
acknowledged for constructing and testing the OASIS ISS
equipment and for providing control of the ISS experiment
during the mission. Drawings of the setup shown here were
reproduced from ZIN design documents. The authors are
grateful to Jennifer Storck and the ZIN engineers for their
dedicated, professional support during the ISS mission. We
are also indebted to Adam Green for helping to monitor
the experiments during this long undertaking. Finally, we
would like to thank the astronauts and cosmonauts for
their enthusiastic engagement and for helping to make
the OASIS mission a success.

\bibliographystyle{elsarticle-num}
\bibliography{smectic_islands.bib}

\end{document}